\documentclass{webofc}
\usepackage[varg]{txfonts}   % Web of Conferences font
\usepackage{subcaption}
\usepackage[font=small,labelfont=bf,textfont=normal,singlelinecheck=false]{caption}

\begin{document}
\title{WLCG Networks: Update on Monitoring and Analytics}
%
% subtitle is optional

\author{\firstname{Marian} \lastname{Babik}\inst{2}\fnsep\thanks{We gratefully acknowledge the National Science Foundation which supported this work through NSF grants \#1148698, \$1836650 and \#1827116.   In addition, we acknowledge our collaborations with the WLCG and LHCONE/LHCOPN communities who also participated in this effort.}
        \firstname{Shawn} \lastname{McKee}\inst{1} \and
        \firstname{Pedro} \lastname{Andrade}\inst{2} \and
        \firstname{Brian Paul} \lastname{Bockelman}\inst{6} \and
        \firstname{Robert} \lastname{Gardner}\inst{4} \and
        \firstname{Edgar Mauricio} \lastname{Fajardo Hernandez}\inst{5} \and
        \firstname{Edoardo} \lastname{Martelli}\inst{2} \and
        \firstname{Ilija} \lastname{Vukotic}\inst{4} \and
        \firstname{Derek} \lastname{Weitzel}\inst{3} \and
        \firstname{Marian} \lastname{Zvada}\inst{3} \\
        for the WLCG Network Throughput Working Group 
        % etc.
}

\institute{ Physics Department, University of Michigan, Ann Arbor, MI, USA
\and
            European Organisation for Nuclear Research (CERN), Geneva, Switzerland
\and 
            University of Nebraska -- Lincoln, Lincoln, NE, USA 
\and       
            Enrico Fermi Institute, University of Chicago, Chicago, IL, USA
\and    
            University of California San Diego, La Jolla, CA, USA 
\and
            Morgridge Institute of Research, Madison, WI, USA
          }
         
\abstract{%
WLCG relies on the network as a critical part of its infrastructure and therefore needs to guarantee effective network usage and prompt detection and resolution of any network issues including connection failures, congestion and traffic routing. The OSG Networking Area, in partnership with WLCG, is focused on being the primary source of networking information for its partners and constituents. It was established to ensure sites and experiments can better understand and fix networking issues, while providing an analytics platform that aggregates network monitoring data with higher level workload and data transfer services. This has been facilitated by the global network of the perfSONAR instances that have been commissioned and are operated in collaboration with WLCG Network Throughput Working Group. An additional important update is the inclusion of the newly funded NSF project SAND (Service Analytics and Network Diagnosis) which is focusing on network analytics. This paper describes the current state of the network measurement and analytics platform and summarises the activities taken by the working group and our collaborators.
%ocusing mainly on the throughput issues that have been reported and resolved during the recent period with the help of the perfSONAR network. We will also cover the updates on the higher level services that were developed to help bring perfSONAR network to its full potential. 
This includes the progress being made in providing higher level analytics, alerting and alarming from the rich set of network metrics we are gathering.  
%inally, we will discuss and propose potential R&D areas related to improving the network throughput in general as well as prepare the infrastructure for the foreseen major changes in the way network will be provisioned and operated in the future.
}
\maketitle
\enlargethispage*{2mm}
\section{Introduction}
\label{intro}
The Open Science Grid (OSG) and the Wordwide LHC Computing Grid (WLCG) have been supporting network monitoring activities since 2012, focusing on assisting their  users and affiliates on improving their overall network throughput by introducing active monitoring of their networks and providing the ability to test for and identify potential network performance bottlenecks \cite{osg,wlcg}. Two important areas of development that were undertaken were establishing and operating a global network of  measurements agents and development and operations of a comprehensive networking monitoring platform, which collects and stores the measurements while making them available for further processing. This has been complemented by several activities that have improved our ability to manage and use both network topology and network metrics to extract clearer understanding of our network problems, locations and bottlenecks via analytics\cite{wlcg-NTW}. 

WLCG Network Throughput Working Group was established in 2014 to help with some of the underlying tasks, such as overseeing the global network of measurement agents based on perfSONAR\cite{ps}, establishing baseline measurements and performing low-level debugging activities. This has lead to a dedicated network throughput support unit, which has proven to successfully coordinate and resolve complex network performance incidents within LHCOPN and LHCONE\cite{lhcone}. 

\begin{figure}[t]
\centering
\includegraphics[width=0.9\textwidth,height=0.9\textheight,keepaspectratio]{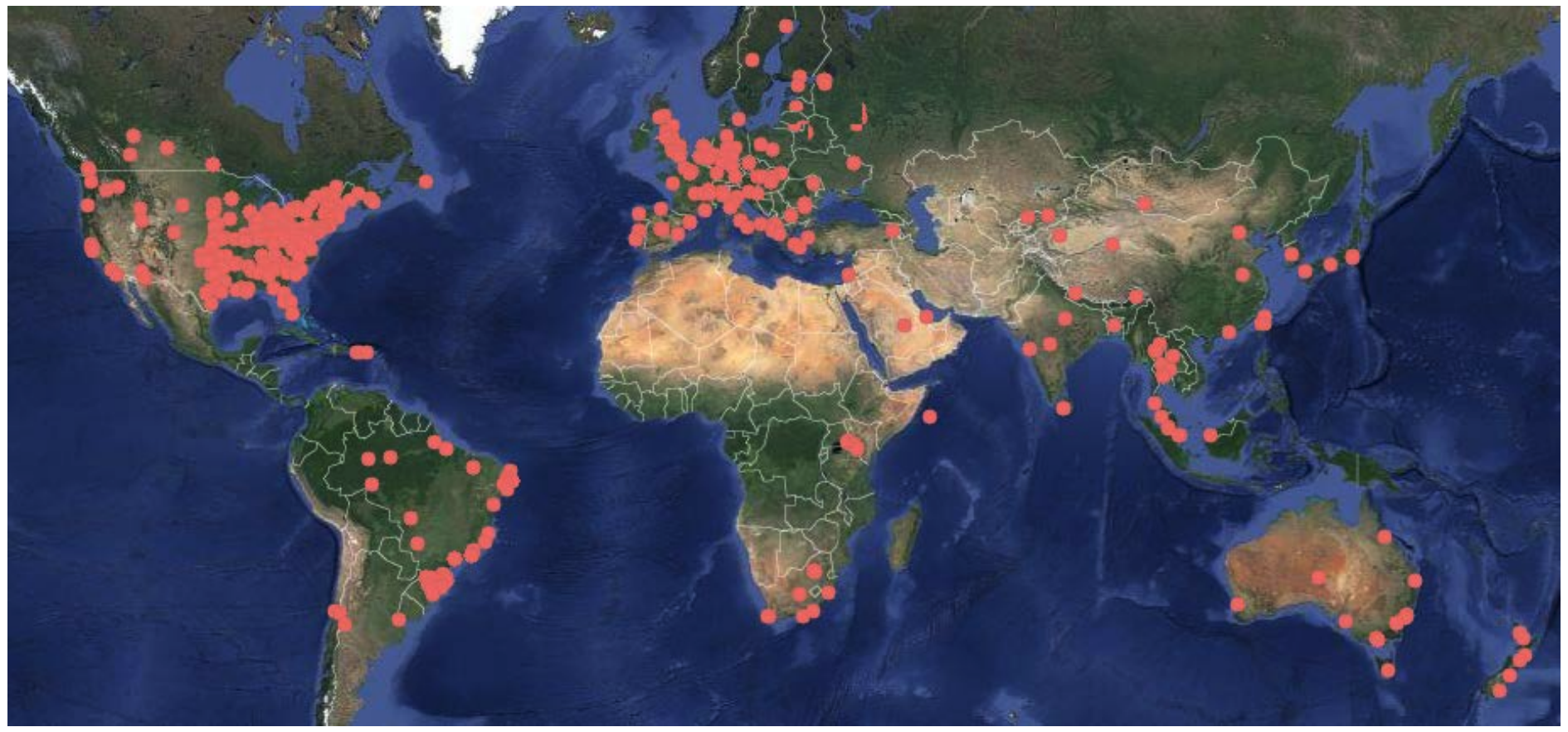}
\caption{perfSONAR public network; there are currently around 2000 known deployed instances with likely an equal number of private deployments, WLCG network is one of the biggest private deployments with over 250 instances connected to LHCOPN/LHCONE.}
\label{fig-1}       % Give a unique label
\vspace*{-15pt}
\end{figure}

\enlargethispage*{2mm}

\section{Network Performance}
Networks that connect sites and experiments need to handle ever increasing amounts of data and convey it across multiple networks around the world. Due to the underlying complexity, end-to-end performance depends on a number of components and their operational status anywhere within the network. When a network is under-performing or errors occur, it can become very difficult to identify and correct the source of the problem as local testing will often not find the cause, as errors can occur anywhere along the path of data as it moves between multiple networks. While disconnect failures are relatively easy to detect and fix, soft failures where a network continues to function but has compromised performance can be very hard to detect. Identification of such problems is best served by the active end-to-end measurements against a predefined target, which in the scope of WLCG and OSG means a global network of agents testing all possible network paths end to end. 

\section{OSG/WLCG Network Monitoring Platform}

Such global network of agents has been established in collaboration with WLCG and OSG sites based on perfSONAR, which is a network measurement toolkit designed to provide federated coverage of paths that helps to establish the end-to-end usage expectations (see Fig. \ref{fig-1}). perfSONAR is open source software, developed by a consortium of ESnet, Internet2, Indiana University, University of Michigan and GEANT\cite{ps}. It provides a number of tools that can take various different network measurements covering different aspects of network functions, bundled in a comprehensive package including tools, scheduler, visualisation and centralised management functions such as configuration and discovery. 

\label{sec-1}
\begin{figure}[th]
\centering
\includegraphics[width=\textwidth]{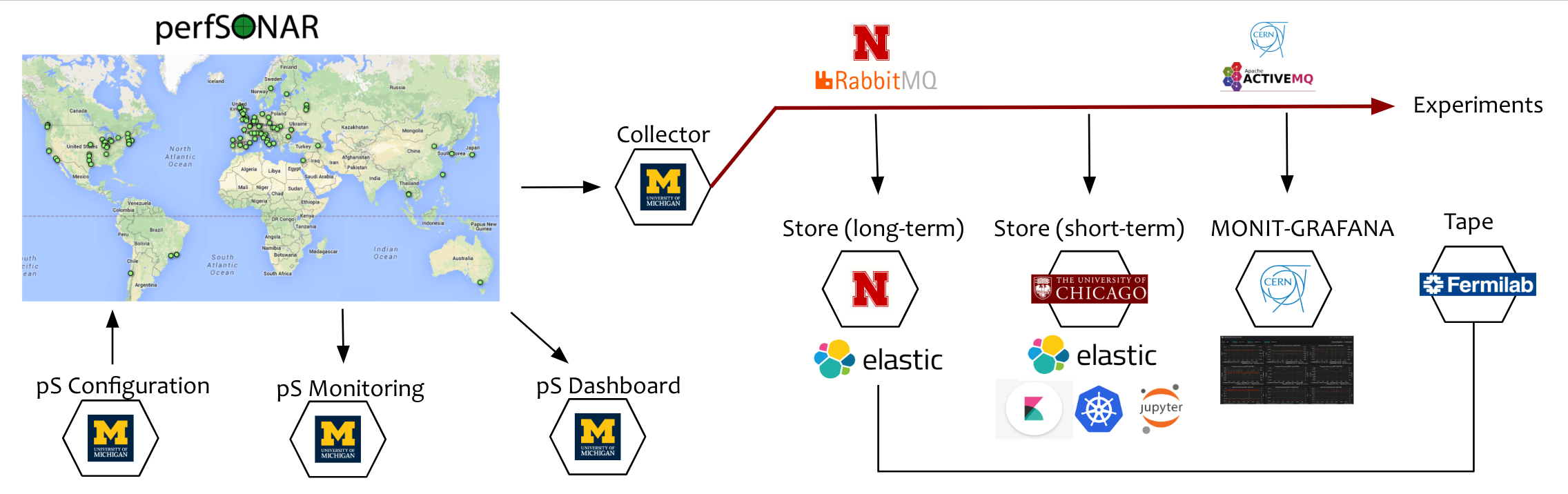}
\caption{OSG Network Monitoring Platform - distributed deployment that collects, stores, visualises and provides APIs for the measurements collected by the WLCG perfSONAR infrastructure} 
\label{fig-4}       % Give a unique label
\vspace*{-15pt}
\end{figure}

The toolkit supports a range of standard metrics that provide useful insights into the current state of the network. For latency and loss, apart from \emph{ping}, it offers implementation of the \emph{one-way and two-way active measurement protocols} (OWAMP/TWAMP)\cite{rfc4656}. An important metric for end-to-end network performance is throughput, which can be measured by three different tools: \emph{iperf3, iperf2 and nuttcp}. The most common is \emph{iperf3}, which can perform memory to memory tests over UDP or TCP and reports TCP retransmits and size of congestion window, which are both very useful in troubleshooting. The final part of the network characteristics is the network path, which can be measured by \emph{traceroute} or \emph{tracepath}, the latter being preferred due path MTU discovery as it can determine \emph{maximum transmission unit} (MTU) along the path and serves as an important indicator of MTU issues which have become quite common.

% importance of measuring our networks

%\begin{figure}[t]
%\centering
%\begin{subfigure}{.45\textwidth}
%  %\includegraphics[width=.4\linewidth]{ps-arch.png}
%  \includegraphics[width=\textwidth]{ps-arch.png}
%  \caption{PerfSONAR Toolkit architecture highlighting main functional areas: tools, scheduling, %archiving, visualisation, configuration and discovery}
%  \label{fig-1}
%  \vspace{-65pt}
%\end{subfigure}%
%\hspace{10pt}
%\begin{subfigure}{.45\textwidth}
%  \includegraphics[width=\textwidth]{owamp.png}
%  \caption{OWAMP - active measurements of delay, loss, ordering, jitter from a pre-defined schedule of UDP packet exchanges.}
% \label{fig-2}
% \vspace{-10pt}
%end{subfigure}%
%caption{(a) PerfSONAR architecture (b) One-way Delay Measurement Protocol (OWAMP).}
%vspace*{-15pt}
%end{figure}

\enlargethispage*{2mm}

OSG has developed and deployed a comprehensive network monitoring platform\cite{osg-datastore} that collects, stores, visualises and further processes all the measurements taken by the perfSONAR infrastructure, see Fig \ref{fig-4}. At its core is a collector, which regularly connects to the remote perfSONAR toolkits, downloads all recent measurements and publishes them to the message bus based on RabbitMQ. This stream is then used to feed three different types of stores, a short-term store located at University of Chicago, which stores data for the last 6 months, a long-term store located at University of Nebraska, which stores the entire dataset and finally a tape system at FNAL, which is used as a persistent backup. The measurements stream is also available to the experiments via ActiveMQ bus at CERN which is populated by a dedicated bridge connected directly to RabbitMQ. The platform is also integrated with the ATLAS Analytics and Machine Learning Platform\cite{atlas_analytics} that makes it easy to combine and analyze network measurements with metrics from various different sources (including Panda, FTS, Rucio, etc.).

The platform also contains a centralised configuration system\cite{PSCONFIG} built upon PWA\cite{PWA}, which is used to configure the tests specifications (tools/measurements specs), meshes (collection of hosts participating in the tests) as well as test schedule for the entire infrastructure. There is also infrastructure monitoring\cite{etf,psetf} that oversees the status of the platform and measurement infrastructure and a set of MaDDash dashboards that visualize the measurement results\cite{psmad}. In addition, there are number of additional dashboards and visualisations available that are discussed in Section 5.

\subsection{Job Network Measurements}
In addition to the metrics collected by perfSONAR, the OSG also collects network metrics from submit hosts within the OSG.  These submit hosts measure the network conditions between the worker nodes and submit hosts during file transfers.  File transfers generally only occur when the job starts and when it completes.  Therefore, the measurements do not capture the status of the connection during job execution.

These job network measurements can capture aspects of the end-to-end path that might be untested by perfSONAR.  For example, in the OSG, worker nodes can be behind a firewall or a NAT device and, in such cases, perfSONAR would often be connected at the network edge and would not be measuring the same network path.

HTCondor is configured to output TCP statistics for data transfer connections between the submit host and the worker node.  The statistics include the number of loss packets, bytes transferred and TCP reordering events.  These statistics are written to a log by HTCondor which is parsed and uploaded by Filebeats\cite{filebeats} in the same datastore we use for perfSONAR metrics.  The data components are parsed and annotated, e.g., we augment transfer records with GeoIP information.

\begin{figure}[th]
\centering
\includegraphics[width=0.9\textwidth,height=0.9\textheight,keepaspectratio]{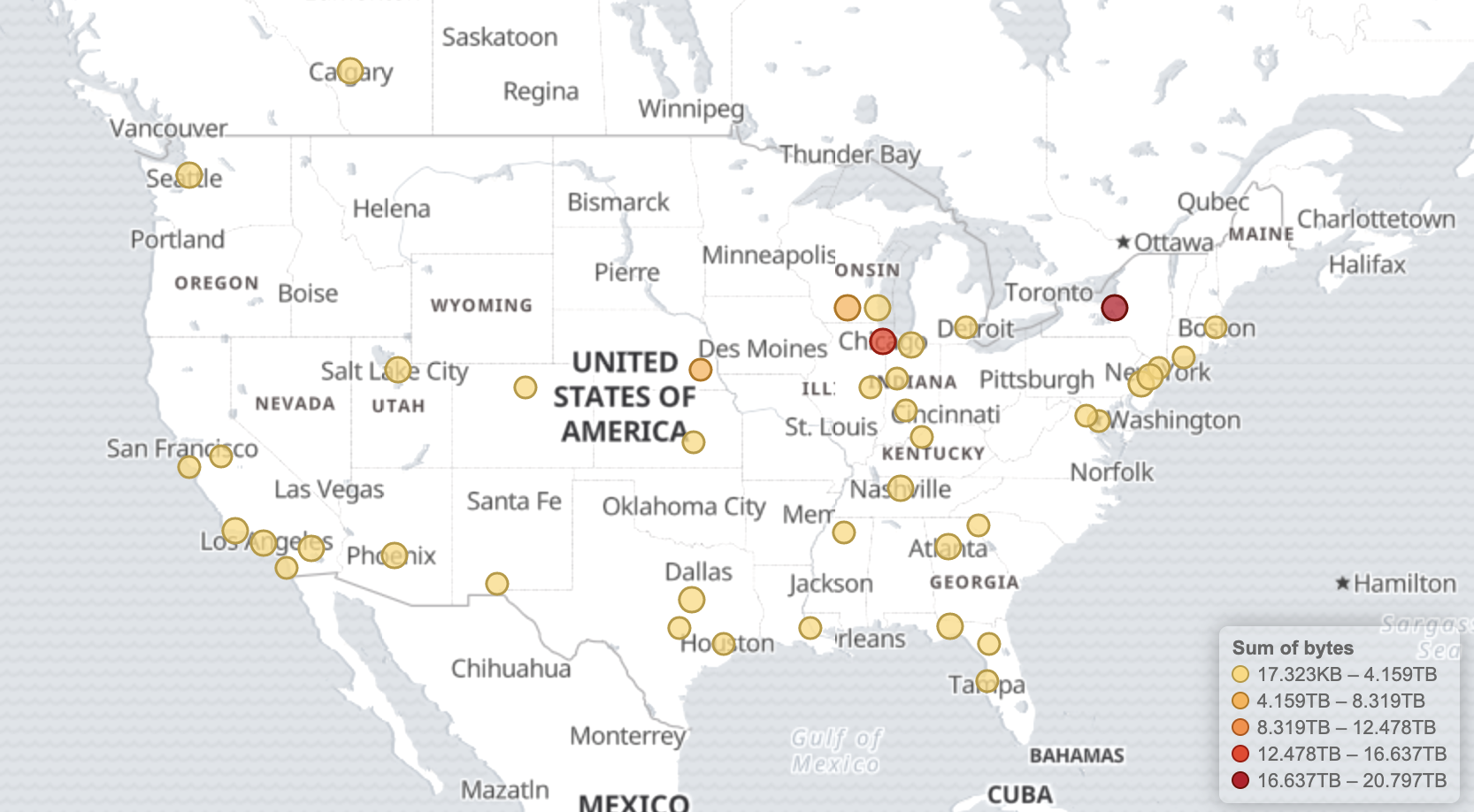}
\caption{Map of destination by bytes transferred to jobs.} 
\label{fig:jobnetwork}       % Give a unique label
\vspace*{-19pt}
\end{figure}

We are just beginning to collect and analyze the job network measurements.  Figure \ref{fig:jobnetwork} shows the data transfer volume to job destinations within the U.S. for January 2020.

\section{Platform Use}
The platform and measurement infrastructure have been used in number of activities and collaborations, improving our understanding of the networks and contributing to the technical evolution and design.

Establishing end-site network throughput support has helped to resolve number of challenging cases that would otherwise be very difficult to detect and isolate or would take considerable amount of time to resolution\cite{net_cases}. In addition, the unit has helped sites with their data centre network design, consulting on the potential bottlenecks caused by the network equipment with insufficient buffers as well as helping to test and benchmark their performance. The feedback gathered from the support unit on the different cases has lead to a discussion and a concrete proposal for MTU recommendations for LHCOPN/LHCONE\cite{mtu_doc}, which aims to improve the overall throughput and standardise MTU deployment across R\&Es and sites. 

There were number of significant contributions to the development and design of network performance monitoring over the years, a notable example is the the current configuration system, which was initially developed as an internal OSG tool and was later adopted by the perfSONAR consortium. Another area of close collaboration was deployment and testing of the IPv6 readiness, which was lead by the HEPiX IPv6 working group\cite{ipv6}. This was a particular example how the platform can be useful in the future to evaluate potential deployment of the new technologies (such as new TCP congestion control algorithms, software defined networks, etc.). Another such example is a collaboration with HELIX NEBULA Science Cloud project, which used the platform to assess network performance of the cloud providers. Finally, close collaborations were established with other research domains and institutes that have also shown interest in network performance and deployment of a similar platform as the one deployed for OSG/WLCG. 

\begin{figure}[th]
\centering
\includegraphics[width=0.9\textwidth,height=0.9\textheight,keepaspectratio]{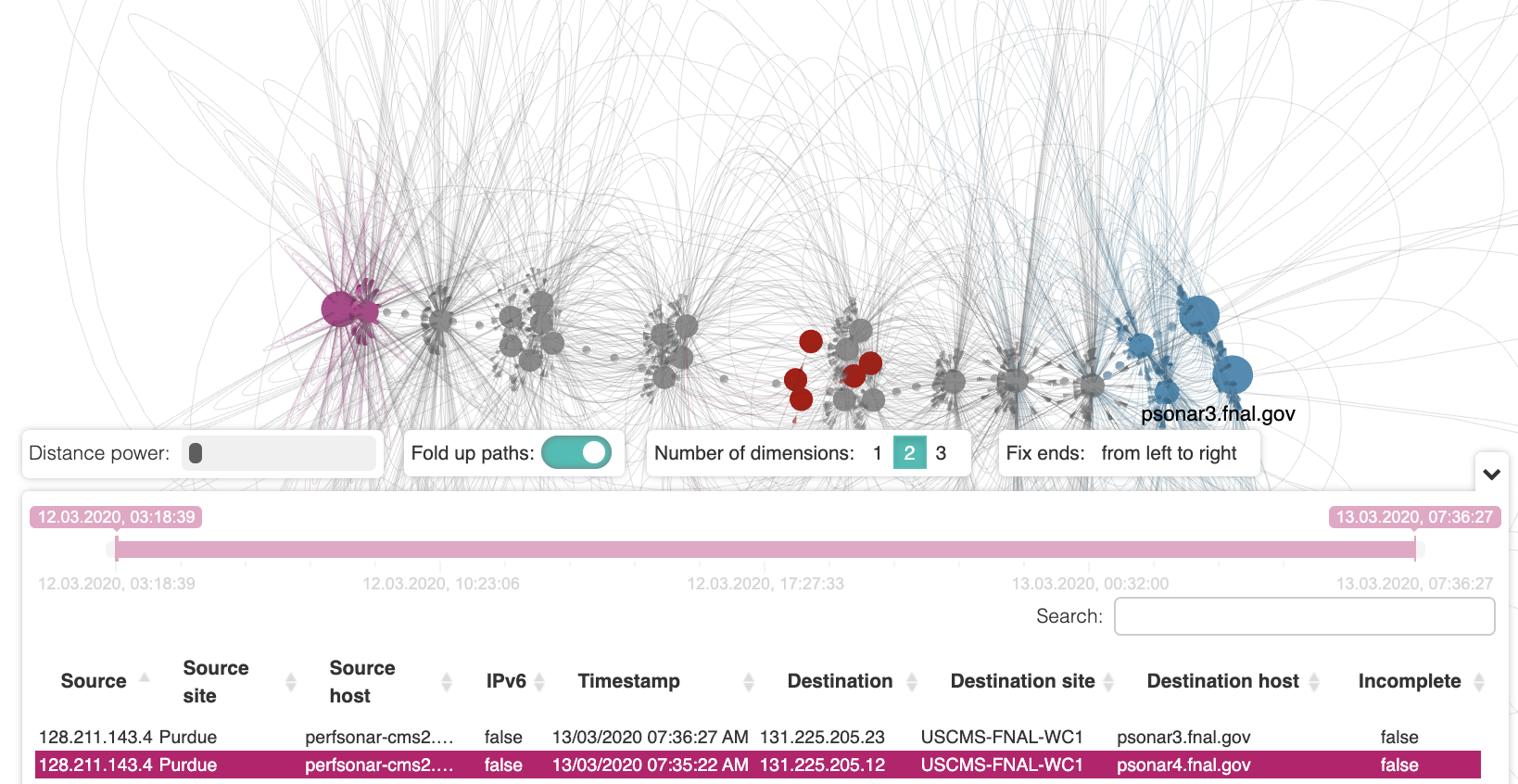}
\caption{Visualisation of multiple network paths as measured by the 
\emph{traceroute} tool between Purdue and Fermi National Accelerator Laboratory.} 
\label{fig:trace}       % Give a unique label
\vspace*{-15pt}
\end{figure}

\section{Network Analytics}

Establishing the \emph{OSG Network Monitoring Platform} and making the data available for experiments and network researchers has triggered great interest from different communities that have started to look at the existing measurements and performed analysis with various different goals. At the same time, the platform has made it possible to diagnose and debug existing network issues, identify the problematic links or equipment and help fix the underlying problems. Among the several past and present projects, the following have delivered notable results or identified important areas where further research is needed: 
\begin{itemize}
\item Real-time detection of “obvious” issues and corresponding altering and notifications have been developed at University of Chicago and is currently being tested as part of the ATLAS Analytics and Machine Learning Platform\cite{atlas_analytics}.
\item A study to derive how LHCOPN network paths perform from the existing OWAMP measurements has shown that OWAMP is sufficiently sensitive to pinpoint when network equipment gets stressed and could be used to easily detect peak periods. The main challenge that still remains is how to extend the model to LHCONE, mainly due to the lack of reliable network traffic data that could be used to train the neural network\cite{babik_borras}. 
\item New visualisation platform for network paths was developed in collaboration with MEPhI\footnote{https://eng.mephi.ru/}, which allows to select and visualise existing paths between two endpoints, see Fig. \ref{fig:trace} 
\item Network path analysis project is currently ongoing at University of Michigan and aims to calculate simple statistics from the existing path measurements in order to auto-detect potential routing problems and help with the visualisation of the measurements.
\item In collaboration with the SAND project\cite{SAND-grant,SAND} and some of the other activities mentioned in this section, we are developing a range of dashboards\cite{NetDashboards} using Kibana to provide distinct insights into the perfSONAR metrics hosted in Elasticsearch.  
\item Understanding the differences between network utilization as seen by R\&E networks as computed from the experiments data transfers is another area of interest. While there has been significant effort contributed to understand network utilisation from the bulk data transfers, there are still major gaps in getting reliable sources of information directly from the R\&E networks.
\end{itemize}

Further analytical studies are planned to better understand our use of networks and how it could be improved. The new versions of perfSONAR plan to integrate direct publishing of the results and configurations needed to operate it globally that would help us make progress in number of areas requiring access to real-time data as well as providing automated debugging and optimisations.

\section{Evolution and Future}
In summary, OSG in collaboration with WLCG have established a comprehensive network monitoring platform that has been used in a number of activities ranging from operations and support and technological deployments up to the research and developments for the network analytics. We have established and made progress in several areas of the
network monitoring and plan to continue to evolve in the same areas also in the near term. There are number of areas where significant R\&D effort will be needed to progress on some of the previously mentioned challenges, but there are also number of opportunities that could provide funding and effort to continue the work. Two projects that will lead the operations and development in the HEP network monitoring are NSF funded IRIS-HEP and SAND. IRIS-HEP will fund the LHC part of Open Science Grid, including the networking area and will create a new integration path (the Scalable Systems Laboratory) to deliver its R\&D activities into the distributed and scientific production infrastructures.  Service Analysis and Network Diagnosis (SAND) will be focusing on combining, visualising, and analyzing disparate
network monitoring and service logging data. It will extend and augment the OSG networking efforts with a primary goal of extracting useful insights and metrics from the wealth of network data being gathered from perfSONAR, FTS, R\&E network flows and related network information from HTCondor and others.

%\newpage
\section{Acknowledgements}
\enlargethispage*{4mm}
We gratefully acknowledge the National Science Foundation which supported this work through NSF grants OAC-1836650 and OAC-1827116.   In addition, we acknowledge our collaborations with the CERN IT, WLCG and LHCONE/LHCOPN communities who also participated in this effort.
%
% BibTeX or Biber users please use (the style is already called in the class, ensure that the "woc.bst" style is in your local directory)
\bibliography{bibliography}

\begin{thebibliography}{21}

\bibitem{osg}
R.~Pordes, D.~Petravick, B.~Kramer, D.~Olson, M.~Livny, A.~Roy, P.~Avery,
  K.~Blackburn, T.~Wenaus, F.~Würthwein et~al. (2007), Vol.~78, p. 012057,
  \urlstyle{tt}\url{http://stacks.iop.org/1742-6596/78/i=1/a=012057}

\bibitem{wlcg}
I.~Bird, P.~Buncic, F.~Carminati, M.~Cattaneo, P.~Clarke, I.~Fisk, M.~Girone,
  J.~Harvey, B.~Kersevan, P.~Mato et~al., Tech. Rep. CERN-LHCC-2014-014.
  LCG-TDR-002 (2014), \urlstyle{tt}\url{http://cds.cern.ch/record/1695401}

\bibitem{wlcg-NTW}
S.~McKee, M.~Babik, S.~Campana, A.D. Girolamo, T.~Wildish, J.~Closier,
  S.~Roiser, C.~Grigoras, I.~Vukotic, M.~Salichos et~al., \emph{Integrating
  network and transfer metrics to optimize transfer efficiency and experiment
  workflows} (2015), Vol. 664, p. 052003,
  \urlstyle{tt}\url{http://stacks.iop.org/1742-6596/664/i=5/a=052003}

\bibitem{ps}
A.~Hanemann, J.W. Boote, E.L. Boyd, J.~Durand, L.~Kudarimoti, R.~{\L}apacz,
  D.M. Swany, S.~Trocha, J.~Zurawski, \emph{PerfSONAR: A Service Oriented
  Architecture for Multi-domain Network Monitoring}, in \emph{Service-Oriented
  Computing - ICSOC 2005}, edited by B.~Benatallah, F.~Casati, P.~Traverso
  (Springer Berlin Heidelberg, Berlin, Heidelberg, 2005), pp. 241--254, ISBN
  978-3-540-32294-8

\bibitem{lhcone}
E.~Martelli, S.~Stancu, \emph{LHCOPN and LHCONE: Status and Future Evolution}
  (2015), Vol. 664, p. 052025,
  \urlstyle{tt}\url{http://stacks.iop.org/1742-6596/664/i=5/a=052025}

\bibitem{rfc4656}
S.~Shalunov, B.~Teitelbaum, A.~Karp, J.~Boote, M.~Zekauskas, RFC 4656, RFC
  Editor (2006)

\bibitem{osg-datastore}
R.~Quick, M.~Babik, E.M. Fajardo, K.~Gross, S.~Hayashi, M.~Krenz, T.~Lee,
  S.~McKee, C.~Pipes, S.~Teige, Journal of Physics: Conference Series
  \textbf{898}, 082044 (2017)

\bibitem{atlas_analytics}
I.~Vukotic, D.~Barberis, F.~Legger, R.~Gardner (ATLAS Collaboration),
  \emph{ATLAS Analytics and Machine Learning Platforms} (2018)

\bibitem{PSCONFIG}
S.~McKee, M.~Babik, (June 2020), \emph{Osg/wlcg psconfig server}, Retrieved
  from \urlstyle{tt}\url{https://psconfig.opensciencegrid.org/}

\bibitem{PWA}
perfSONAR Developers, (June, 2020), \emph{{pSConfig Web Admin}}, Retrieved from
  \urlstyle{tt}\url{https://docs.perfsonar.net/pwa.html}

\bibitem{etf}
M.~Babik, (June 2020), \emph{Experiments {Test} {Framework} ({ETF})}, Retrieved
  from \urlstyle{tt}\url{https://etf.cern.ch/docs}

\bibitem{psetf}
M.~Babik, (2019), \emph{{perfSONAR ETF Monitoring}}, Retrieved from
  \urlstyle{tt}\url{http://etf.cern.ch/docs/latest/user/overview.html#service}

\bibitem{psmad}
perfSONAR Consortium, (June 2020), \emph{{perfSONAR} {Monitoring and Debugging
  Dashboard (MADDASH)}}, Retrieved from
  \urlstyle{tt}\url{http://psmad.opensciencegrid.org/maddash-webui/index.cgi}

\bibitem{filebeats}
E.~Inc, (June 2020), \emph{Filebeat: Lightweight log analysis \&
  elasticsearch}, Retrieved from
  \urlstyle{tt}\url{https://www.elastic.co/beats/filebeat}

\bibitem{net_cases}
M.~Babik, (June 2020), \emph{Network throughput support unit documentation},
  Retrieved from
  \urlstyle{tt}\url{https://twiki.cern.ch/twiki/bin/view/LCG/NetworkTransferMetrics#Network_Throughput_Support_Unit}

\bibitem{mtu_doc}
S.~McKee, M.~O'Connor, (June 2018), \emph{Jumbo frame considerations for
  {LHCOPN/LHCONE}}, Retrieved from
  \urlstyle{tt}\url{https://indico.cern.ch/event/764495/}

\bibitem{ipv6}
S.~Campana, K.~Chadwick, G.~Chen, J.~Chudoba, P.~Clarke, M.~Eliáš, A.~Elwell,
  S.~Fayer, T.~Finnern, L.~Goossens et~al., \emph{WLCG and IPv6 – the HEPiX
  IPv6 working group} (2014), Vol. 513, p. 062026,
  \urlstyle{tt}\url{http://stacks.iop.org/1742-6596/513/i=6/a=062026}

\bibitem{babik_borras}
M.~Babik, H.~Borras, Tech. Rep. CERN-IT-Note-2017-001, CERN, Geneva (2017),
  \urlstyle{tt}\url{http://cds.cern.ch/record/2252410}

\bibitem{SAND-grant}
B.~Bockelman, (2018), \emph{Cc* integration: Service analysis and network
  diagnosis (sand)}, Retrieved from
  \urlstyle{tt}\url{https://www.nsf.gov/awardsearch/showAward?AWD_ID=1827116}

\bibitem{SAND}
D.~Weitzel, (June, 2020), \emph{Website for service analysis and network
  diagnosis project}, Retrieved from \urlstyle{tt}\url{https://sand-ci.org/}

\bibitem{NetDashboards}
S.~McKee, (2020), \emph{Prototype {OSG/WLCG} kibana network dashboards},
  online, Retrieved from
  \urlstyle{tt}\url{https://atlas-kibana.mwt2.org/s/networking/app/kibana#/dashboard/07a03a80-beda-11e9-96c8-d543436ab024?_g=()}

\end{thebibliography}

\end{document}